# Physicochemical characterization of low rank Nigerian coals

Bemgba Bevan Nyakuma

*Centre for Hydrogen Energy, Institute of Future Energy,
Universiti Teknologi Malaysia, 81310 Skudai,
Johor Bahru, Malaysia.*

Email: bbnyax1@gmail.com, bnbevan2@live.utm.my

**Abstract**

This study is aimed at investigating the physicochemical properties of low rank Nigerian coals. The coal samples were examined using bomb calorimetry, ultimate and proximate analyses to determine the heating value, elemental composition and physicochemical properties. The calorific values of the coals were; 23.74 MJ/kg for GMG; 27.34 MJ/kg for SKJ and 30.52 MJ/kg for AFZ. The results of elemental analysis showed high C, H, O and low N, and S content for the coals. The GMG coal displayed the lowest N, and S values emphasizing its low pollutant emission potential. The highest VM (51.16 %) and M (5.28 %) content was observed for GMG; AFZ coal displayed the highest ash content at 30.99 %; and SKJ the highest FC content. Consequently, the reactivity and maturity of the low rank Nigerian coals investigated increases in the order; GMG > SKJ > AFZ.

## Introduction

Coal is considered the cheapest and most widely abundant fossil fuel in the world (Kurt, 2008; Heinberg and Fridley, 2010; Ye *et al.*, 2013). According to the United States Energy Information Administration (EIA), the cost of generating 1 million British Thermal Units (Btu) of energy from coal is $1.69 against $6.94 and $6.23 for natural gas and petroleum, respectively (EIA, 2008). By current estimates, coal consumption accounts for over 60 % of global fossils fuels utilization compared to 17 % and 19 % for natural gas and oil, respectively (OECD, 2012). Consequently, coal has become the largest source of primary fossil fuel energy utilization for electric power generation and feedstock for chemicals, fuels and steel production (Buhre *et al.*, 2005; Minchener, 2005; Ye *et al.*, 2013; Sarwar *et al.*, 2014). At present, the most practical technologies for coal conversion and utilization are carbonization, gasification and combustion (Buhre *et al.*, 2005; Speight, 2012; Ni *et al.*, 2014). Although, these underlying dynamics have fuelled the global consumption of coal, energy self-sufficiency predominantly in the emerging economies of sub-Saharan Africa remains a colossal conundrum.

Nigeria is the largest economy and exporter of crude oil in Africa with estimated crude oil reserves of 35 billion barrels, 187 trillion cubic feet of natural gas, and over 4 billion metric tonnes of coal. Furthermore, the efficacious exploration for coal has led to the discovery of vast coal deposits in 14 of the 36 states of the country (Ohimain, 2014). In spite of Nigeria's vast energy potential, the country remains perennially plagued by an energy crises (Ibitoye and Adenikinju, 2007). This has resulted in low electric power generation, poor distribution and transmission losses and perpetual blackouts. This unfortunate scenario have greatly undermined Nigeria's potential for sustained socioeconomic growth, infrastructural development and energy security. Therefore, there is a critical need for low-cost, sustained electricity production and power supply, generated from cheap sources of energy such as coal. With cheap coal based electric power, the real sector of Nigerian economy will be greatly enhanced, spurring socioeconomic growth, infrastructural development, jobs creation and improved living standards.

However coal is a "dirty fuel" which accounts for 37 % of global carbon dioxide emissions. Hence concerns about pollutant emissions and environmental sustainability make coal utilization for power generation a controversial topic especially in developing countries. The Intergovernmental Panel on Climate Change (IPCC) warns that an 80 % reduction of GHGs emissions is required to keep temperatures below 2 °C (Stocker *et al.*, 2013). In view of this, efficient low-carbon emission techniques such as clean coal technologies, underground coal gasification (UCG), carbon capture and storage and Integrated Gasification Combined Cycle (IGCC) are currently under investigation for future implementation in coal power generation (Khadse *et al.*, 2007; Li and Fan, 2008; Franco and Diaz, 2009; Sheng *et al.*, 2014; Khadse, 2015; Sanchez *et al.*, 2015). However, the effective implementation of the efficient low-carbon emission techniques for coal power generation requires comprehensive knowledge of the physical and thermochemical properties of the feedstock coals.

Parenthetically, the lack of scientific data on the physical and thermochemical properties of Nigerian coals has greatly hampered its utilization for various electric power and industrial applications. This is vital in examining the classification (ranking) and feasibility of coal resources as potential feedstock (Björkman, 2001). The most common techniques for examining the properties of coals include elemental composition, proximate analysis and calorific heating value. Furthermore, this data is essential for engineering design, process optimization and project costing of coal conversion systems. Consequently, this paper presents the physicochemical properties of coal samples from three newly discovered deposits in Edo, Gombe and Nasarawa states of Nigeria. The study examines the rank and reactivity of the coals and proffers potential future applications.

**Experimental**

**Materials**

Coals samples were obtained from three different locations; Afuze coalfields in Afuze, Edo state; Shankodi-Jangwa coal seam in Obi, Nasarawa state; and the Garin Maiganga coal mines in Akko, Gombe state of Nigeria. The coal samples were labelled; AFZ - Afuze coal; SKJ –

Shankodi-Jangwa and GMG – Garin Maiganga. Figure 1 presents the map showing the location of the coals in the sedimentary basins of Nigeria (Obaje, 2009).

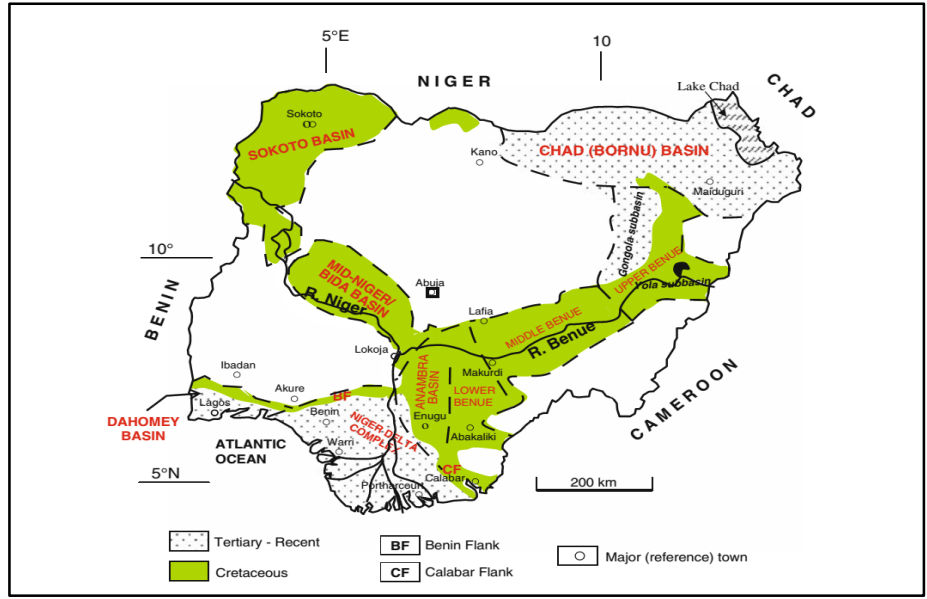

**Figure** 1: Sedimentary basins of Nigeria (Obaje, 2009).

The AFZ coal is reportedly a sub-bituminous coal obtained from Afuze coalfields in Edo state, located south west of the Benin Flank in the Anambra Basin (Equatorial Mining & Exploration, 2015; Ohimain, 2014). The SKJ coal is a bituminous coal acquired from the Shankodi-Jangwa seam in Obi, Nasarawa state which lies in the Benue trough (Jauro *et al.*, 2007; Ryemshak and Jauro, 2013). The GMG coal is a sub-bituminous coal obtained from the Garin Maiganga coal mines in Akko, Gombe state of Nigeria (Ryemshak and Jauro, 2013).

**Methods**

The coal samples were crushed and sifted using a 250 μm mesh Retsch™ analytical sieve to obtain homogeneous sized particles for physicochemical characterization. Next, the elemental composition of the coal samples was carried out using elemental (CHNS/O) analyser (EL Vario MICRO Cube, Elementar™) according to ASTM D5291 standard technique. Proximate analysis was examined using American Society for Testing and Materials (ASTM) D3173, D3174 and D3175 standard techniques for moisture, volatile matter and ash content, respectively, while fixed carbon was determined by difference. The mineral matter was calculated from the Parr formula (Mm = $1.08A + 0.55S$); where *A* and *S* represent Ash and Sulphur content, respectively. Next, the gross calorific or higher heating value (in MJ/kg) of the coals was determined using bomb calorimetry (IKA C2000 Bomb Calorimeter). All tests were repeated at least three times to ensure the reliability of the results.

**Results**

**Ultimate Analysis**

The results for the elemental composition and the calorific value of the coals are presented in Table 1 with all values reported in as received basis (a.r).

Table 1: Ultimate analysis of low rank Nigerian Coals.

| Property | Symbol | GMG (wt. %) | SKJ (wt. %) | AFZ (wt. %) |
|---|---|---|---|---|
| Carbon | C | 61.96 | 71.46 | 72.46 |
| Hydrogen | H | 4.42 | 6.40 | 6.07 |
| Nitrogen | N | 1.07 | 1.37 | 1.63 |
| Sulphur | S | 0.39 | 2.03 | 1.41 |
| Oxygen | O | 32.16 | 18.76 | 18.43 |
| Heating value | HHV | 23.74 | 27.34 | 30.52 |

As can be observed in Table 1, the coal samples possess high C, H, O content, however low values were also obtained for N, and S during elemental analysis. The highest carbon C content was observed in AFZ coal while GMG was lowest. However, the values of C for AFZ and SKJ differ by only 1 wt %. The highest H content was observed in SKJ while the lowest value overall was observed in GMG coal. The C and H content of coals is significantly related to the maturity (rank), calorific value and chemical reactivity during thermal conversion (Speight, 2012; Ryemshak and Jauro, 2013).

The lowest N content was observed in GMG coal whereas the values for SKJ and AFZ were 1.37 wt.% and 1.63 wt.%, respectively. Furthermore, the lowest S content was observed in GMG, whereas the values for SKJ and AFZ were 2.03 wt. % and 1.41 wt. %, respectively. The N and S content is an indication of the environmental friendliness of the fuels relative to potential NOx and SOx pollutant emissions. Hence, the high S content in SKJ and AFZ may limit future applications of the coals particularly in steel manufacturing. Overall, GMG coal showed the lowest N and S content and may be considered the coal with the lowest potential environmental pollutants. According to the study by (Ryemshak and Jauro, 2013) the properties of GMG is suitable for cement and steel manufacture, power generation as well as industrial and domestic heating.

The calorific or heating value of the coals were; GMG - 23.74 MJ/kg; SKJ - 27.34 MJ/kg and 30.52 MJ/kg for AFZ. The results were found to be in good agreement with values for GMG (20.86 MJ/kg) and SKJ (27.22 MJ/kg) reported in literature (Ryemshak and Jauro, 2013). Furthermore, the calorific values for the coals were consistent with the C and O content of the coals. More so, the comparatively lower calorific values of GMG and SKJ can be attributed to the higher values of O presented in Table 1. Consequently, AFZ coal showed the highest value of C content and lowest value of O content which is translated in the highest calorific of the coals investigated. Hence, the results indicate that reactivity and maturity of the coal increases in the order; GMG → SKJ → AFZ.

**Proximate Analysis**

The results of proximate analyses and mineral matter of the coals, reported in as received basis (a.r), are presented in Table 2.

**Table 2**: Proximate analysis of low rank Nigerian Coals.

| Property | Symbol | GMG (%) | SKJ (%) | AFZ (%) |
|---|---|---|---|---|
| Moisture | M | 5.28 | 5.14 | 1.97 |
| Volatile Matter | VM | 51.16 | 40.73 | 45.80 |
| Ash | A | 21.05 | 14.94 | 30.99 |
| Fixed Carbon | FC | 22.52 | 39.18 | 21.24 |
| Mineral matter | Mm | 22.95 | 17.25 | 34.24 |

The results indicate that lowest M content was observed in AFZ coal. Conversely, the M content observably increased in the order; AFZ→SKJ→GMG. Consequently, the results for the M content corroborate the order of maturity of the coals; GMG → SKJ → AFZ inferred for the coals. The M and VM content of coals is an index for evaluating the maturity, quality and potential application of different coals. Furthermore, the classification of coals as either agglomerating or non-agglomerating is based on the determined VM values (Speight, 2012). The highest VM content was observed for GMG, with the values 40.73 % and 45.80 % reported for SKJ and AFZ, respectively. In addition, the results also indicate that SKJ and AFZ are agglomerating coals as against GMG which is non-agglomerating. Overall, the lower the VM content, the higher the ranking or maturity of the coal. Based on the VM criteria, SKJ can be considered the most matured of the coals, followed by AFZ and lastly, GMG.

The ash content, or bulk mineral matter, is used to determine the fouling or slagging potential of coals during thermal conversion (Speight, 2012). Furthermore, ash content reportedly affects the composition, volume and performance of blast furnace coke (Obaje and Ligouis, 1996; Ryemshak and Jauro, 2013). The results of ash analysis in Table 2, indicate that AFZ coal displayed the highest ash content of 30.99 %; compared to 21.05 % for GMG and 14.94 % for SKJ coal. Similar trend was observed for the mineral matter of the coals. Hence, from the results we can reasonably infer that the fouling potential of the coals is in the order; SKJ→GMG→AFZ.

The highest FC content was observed in the SKJ coal followed by GMG and AFZ. Since, FC is the solid residue leftover after devolatization and can be used to estimate the amount of coke obtainable form coal carbonization (Speight, 2012; Ryemshak and Jauro, 2013). Therefore, SKJ has the highest coke potential among the coals investigated in this study. Similar observations have been reported in literature (Ryemshak and Jauro, 2013).

## Conclusion

The physicochemical properties of three low rank Nigerian coals was characterized to determine their elemental composition, proximate analysis, mineral matter and calorific values. The elemental analysis showed that the coal samples possess high C, H, O and low the N, and S values. The properties of GMG are suitable for applications such as domestic heating and power generation; while high coking potential of SKJ is useful for steel manufacture; and AFZ could be used for cement manufacture. Overall, the results indicate that reactivity and maturity of the coals increases in the order; GMG➔ SKJ ➔ AFZ.